**Concentrated sunlight for accelerated stability testing of organic photovoltaic materials: Towards decoupling light intensity and temperature**

I. Visoly-Fisher,[a,b,1] A. Mescheloff,[a] M. Gabay,[c] C. Bounioux,[a] L. Zeiri,[c] M. Sansotera,[c,2] A. E. Goryachev,[a] A. Braun,[a,3] Y. Galagan,[d] E.A. Katz,[a,b*]

[a]Department of Solar Energy and Environmental Physics, J. Blaustein Institutes for Desert Research, Ben-Gurion University of the Negev, Sede Boker Campus 84990, Israel
[b]Ilse Katz Inst. of Nano-Science and Technology, Ben-Gurion University of the Negev, Be'er Sheva 84105, Israel
[c]Department of Chemistry, Ben-Gurion University of the Negev, Be'er Sheva 84105, Israel
[d]Holst Centre, PO Box 8550, 5605 KN Eindhoven, the Netherlands

**Abstract**

We have demonstrated OPV accelerated degradation studies using concentrated sunlight, where the atmosphere, temperature and illumination intensity were *independently* controlled. Testing various schemes for controlling the sample temperature under concentrated sunlight showed that heating of P3HT:PCBM was caused by photons at the absorbed wavelength range and dissipation of excess photon energy, and not necessarily by IR photon absorption. Sunlight chopping was found to be an effective method for independent temperature control under illumination by concentrated sunlight.

The first accelerated degradation tests using sunlight concentration applied to P3HT:PCBM *blends* were reported. P3HT:PCBM blends exposed to concentrated sunlight in the presence of traces of oxygen/ humidity showed degradation induced by photo-oxidation of the P3HT backbone within the P3HT:PCBM blend, which is significantly thermally accelerated, in agreement with previous observations. However, this could be demonstrated in a time scale of minutes and hours, i.e., significantly accelerated. Exposure of well encapsulated P3HT:PCBM films to concentrated sunlight demonstrated stability up to 3,600 sun*hours, corresponding to about 1.6 years of operating time. This result was obtained at 300 suns exposure after merely 12 hours, demonstrating the advantage of using concentrated sunlight for accelerated stability tests. These tests can therefore combine extremely high acceleration factors with profound understanding of the effect of various, independently controlled factors on the degradation mechanisms.

---

[1] Corresponding authors: irisvf@bgu.ac.il, tel: +972-8-6563500, keugene@bgu.ac.il, tel: +972-8-6596739
[2] Present address: Dipartimento di Chimica, Materiali e Ingegneria Chimica "Giulio Natta", Politecnico di Milano, Milan, Italy
[3] Present address: Imperial College London, London SW7 2AZ, UK

## 1. Introduction

The development of organic photovoltaics (OPV) faces the noteworthy challenge of combining high efficiency and operational stability. Recently, the lifetime of polymer solar cells have been considerably improved. For example, well-encapsulated bulk heterojunction solar cells based on poly(3-hexylthiophene) (P3HT) blended with the fullerene derivative [6,6]-phenyl C60 -butyric acid methyl ester (PCBM) [1-4] or poly[9'-hepta-decanyl-2,7-carbazole-alt-5,5-(4',7'-di-2-thienyl-2',1',3'-benzothiadiazole) (PCDTBT) blended with PC70BM [1] have shown lifetimes well beyond a year. This achievement raises the need for relevant accelerated tests of the operational lifetime, especially as a rapid tool for screening newly developed materials and devices. Accelerated testing can be even more crucial for the study of degradation mechanisms and stability of 'small molecule' OPV devices [5-7], as these bilayer devices exhibited longer lifetimes than polymer-based cells [8].

The introduction of various aggressive conditions can increase the degradation rate relative to conventional degradation at standard test conditions (1 sun = 100 mW/cm$^2$, 25$^o$C) thus decrease the test time. Stability testing of polymer solar cells and their components is commonly done under natural or simulated 1 sun illumination, and can be accelerated by elevated temperatures with acceleration factors up to slightly over 20 [9-12]. The use of natural and simulated *concentrated* sunlight has recently shown the potential for further accelerated stability testing of OPV devices [13-17] and materials [14, 16]. Concentrated light degradation testing of non-encapsulated layers of pure polymers utilized in OPV active blends demonstrated an acceleration of the polymer photo-bleaching in air, with acceleration factors for degradation in P3HT absorption exceeding 100 for exposure to 150 suns or more [13, 14, 16].

Degradation mechanisms in OPV are complex and include a variety of processes: photo-bleaching of the conjugated polymers (which in turn can be governed by various mechanisms, e.g., the attack by singlet oxygen, formed by energy transfer from the polymer's triplet state [18-20], or photo-generated radical oxidation [21-23]), photo-oxidation of fullerene moieties [24, 25], degradation induced by the hole conducting poly(3,4-ethylenedioxythiophene) poly(styrenesulfonate) (PEDOT:PSS) layer [26-29], ion migration from the electrodes and degradation at these interfaces [12, 30-34], morphological changes in the device [24, 35-37], etc. Most of these processes are stimulated by external factors such as light, heat, oxygen or moisture [13], and can be affected by the composition and structure of the OPV materials and the device architecture. As a result, they are almost inseparable under standard device testing and operation. Understanding these degradation mechanisms, which is essential for the development of stable OPV materials and devices, requires detailed characterization of the factors affecting them.

Sample temperature rise due to exposure to concentrated light is a fundamental problem for OPV accelerated degradation testing, which becomes more problematic with increased light intensities and acceleration factors. Such temperature rise can cause difficulties in analyzing the outcome of concentrated light testing, including

- nonlinear dependence of the acceleration factor on light intensity [14], which complicates comparison of the results obtained at various light intensities even for the same exposure dose (1 *sun·hour* has been suggested as a dose unit [14]), and
- difficulties in separating light induced mechanisms from those controlled by the sample temperature.

Deviation from linear dependence of the degradation rate of P3HT absorption on the light intensity was previously attributed to thermal acceleration of the photo-oxidation caused by sample heating under concentrated light in air. The photo-oxidation mechanism was conserved at various light intensities [16]. Photolysis at nitrogen atmosphere was not thermally activated and was therefore not affected by the temperature [14].

Herein we demonstrate stability testing with highly concentrated natural sunlight up to ~5,000 suns, where the various parameters affecting degradation - atmosphere, temperature and illumination intensity - were *independently* controlled. We report the first accelerated tests using concentrated sunlight applied to P3HT:PCBM *blends* with bulk heterojunction morphology, the 'work-horse' of polymer OPV. Highly concentrated sunlight was used (compared to the concentration levels used for such tests of non-encapsulated layers of pure P3HT [14]) since PCBM is known to increase the P3HT:PCBM blend stability due to excited state quenching [13, 38, 39]. The degradation in light absorption in the OPV photoactive layer (photo-bleaching) was used to quantify the degradation. Various spectroscopic methods were utilized to determine the degradation mechanism(s). The results demonstrate our ability to separate light- from heating- effects on degradation up to very high sunlight concentration, and show the effect of these factors on P3HT degradation mechanisms in the P3HT:PCBM blend - photo-oxidation and photolysis.

**2. Experimental details**

2.1. Sunlight concentration

Outdoor sunlight was concentrated and transferred indoors using two systems, both developed in-house: (1) in the 'Mini-dish' system sunlight is concentrated using a paraboloidal mirror and focused into a transmissive (quartz-core) optical fiber of 1mm in diameter and delivered indoors (Figure S1, Supplemental Information) [40, 41]. The intensity was modulated using a pizza-slice iris. (2) In the 'Solar furnace' system a dual-axis tracking flat heliostat reflects sunlight into the laboratory, where a flat mirror (with a hole at its center) tilted at $45^o$ redirects the light upward to a 526 mm-diameter paraboloidal dish of numerical aperture of 0.4, whose focal plane is just below the tilted mirror (Figure S2) [42, 43]. The light intensity was moderated by a louvered shutter between the heliostat and the flat indoor mirror. It should be noted that the spectrum measured at 'noon time ± 2-3 hours' at Sede Boqer (Lat. $30.8^oN$, Lon. $34.8^oE$, Alt. 475 m), where the lab is located, is very close to the AM 1.5G spectrum.

Flux uniformity in both systems was achieved using kaleidoscopes placed between the distal fiber tip ('Mini-dish') or the paraboloid focal point ('Solar furnace') and the cell (Figure S3). An increase in the kaleidoscope exit area resulted in deconcentration of the sunlight delivered to the cell. The degree of concentration can be varied gradually from 0 to 100 and 10,000 suns for 1 $cm^2$ and 1 $mm^2$ illuminated areas, respectively. The incident power of concentrated sunlight was measured with a pyrometer of 5% accuracy, and the sunlight concentration was calculated taking into account the cell illuminated area. Exposure to 1 sun was performed outdoors with 8 hours/day of direct sunlight irradiation. Sunlight exposure dose was quantified in units of sun$^*$hours (1 sun*hour = 360 $J/cm^2$). The reported doses are the actual dose that the samples were exposed to; in the case of chopped illumination only the illuminated period in each cycle was taken into account.

2.2 Sample temperature control

The sample was thermally bonded to the top of a thermoelectric or water cooled plate. The temperature was set to $5^oC$ unless otherwise noted. To further reduce heating, a 16.5×16.5 $cm^2$, 2.5 mm thick absorbing short-pass filter (K-5, Schott) was introduced to the 'Solar furnace' to filter IR radiation (wavelengths longer than 900 nm, Figure S4). A sidelong fan was set to prevent overheating of the filter during irradiation. While the absorption of P3HT:PCBM is limited to wavelengths shorter than 675 nm, longer wavelength radiation might be expected to increase the cell's temperature if absorbed elsewhere (glass, back contact etc.).

The 'Solar furnace' can operate in continuous-irradiation or flash-like mode. The latter is achieved by inserting a reflective rotating disk (chopper) of diameter 264 mm with a 30×50 mm aperture close to its circumference for sample illumination, positioned just above the focal plane (Figure S5 a). The disk's rotation frequency can be varied up to 25±1 Hz, and the shortest temporal window for constant sample irradiance is ~1.1 ms. According to the geometry of the chopper disk, the sample is irradiated 5.4 % of the cycle time (Figure S5 b). The light power $P_{in}$ during 'light on' periods is equal to that during continuous-illumination. Chopping the light enables effective dissipation of the heat, accumulated during the 'light on' period, during the non-irradiated period, hence better control of the sample's temperature.

A thermocouple (T type) was connected using silver paste to the rear (glass) side of the samples. Silver paste is an excellent heat conductor with high reflectivity in order to minimize light absorption by thermocouple. The thermocouple was connected to a Campbell data storage device for temperature reading and displaying. Temperature readings were taken after the system reached thermal equilibrium, few seconds up to more than 30 minutes, depending on the irradiation intensity.

## 2.3 Sample preparation

P3HT:PCBM (bulk heterojunction) blend films were prepared using spin casting of the blend solution followed by encapsulation. Microscope glass slides were cut to size 2×2 cm$^2$ and cleaned with a compressed air stream, 5 minutes sonicating in acetone, methanol and 2-propanol, consecutively, dried under compressed air, and exposed to an UV-ozone treatment for 15 minutes (NovaScan PSD). 45 mg of P3HT (Rieke Metals, regioregular electronic grade, average molecular weight: 50-70K) and 45 mg of PCBM (Solenne BV, 99.5 %) were dissolved in 3 ml of dry chlorobenzene (Acros, 99.8%). The solution was steered for one hour at 60ºC under nitrogen, and then spin-coated at 650 RPM (Laurell, WS-650MZ-23NPP/LITE) on a clean glass plate (about 0.2 ml solution per plate). The resulting film thickness was 200+/-10 nm. No annealing was performed as part of the sample preparation. These samples were encapsulated using the 'glass-on-glass' configuration. A stripe of about 2 mm width of the P3HT:PCBM layer was removed and cleaned around the edges of the base plate, using a Q-tip soaked with acetone. 2 mm wide double-sided Scotch® tape stripes were cut and placed around the edges of the glass substrate and the plates glued together inside the glove box, with the edges sealed with Teflon tape and Parafilm; one on top of the other. Since it does not entirely prevent penetration of ambient species into the sample, but rather slows it, we refer it as 'permeable encapsulation'. On the other hand, this encapsulation was reversible, and could be opened for further investigation of the P3HT:PCBM layer after sunlight exposure.

Samples with high-quality thin film encapsulation were developed and supplied by the Holst Centre [44], with very low water vapor transition rate (WVTP) and oxygen transition rate (OTR) of 10$^{-6}$ g·m$^{-2}$·day$^{-1}$. In these samples, annealed P3HT (Plextronics, Plexcore OS 2100) :PCBM (Solenne BV, 99 %) active layer was encapsulated between a glass substrate and thin-film barrier layer, typically used in ITO-free PPV cells [45]. To provide hermetic sealing, the organic layer at the edges of the samples was removed via laser ablation. The effect of ITO and PEDOT:PSS on the P3HT:PCBM active layer degradation was examined using samples with the layer sequence glass/ITO and/or PEDOT:PSS /P3HT:PCBM/plastic barrier layer (with no back electrode). This type of encapsulation could not be opened, hence these samples were characterized using absorption spectroscopy only. The P3HT:PCBM film thickness in these samples was 200 nm, PEDOT:PSS film thickness was 100 nm.

2.4 P3HT:PCBM characterization

UV-Vis absorption spectra were measured in transmission mode on Jasco V-570 and V-630 UV/VIS/NIR spectrophotometers at the range of 350-700 nm, bandwidth: 1.5nm. Absorption measurements were also performed using the optical modes of a spectral response measurement system ("TECHNOEXAN" Ltd, Ioffe Physical-Technical Institute, Russia) in the range 300-700 nm. The absorption degradation state was calculated as the percent ratio of the total number of photons absorbed after sunlight exposure to that absorbed before exposure. The number of photons absorbed at each wavelength was calculated using the measured absorbance spectrum and the number of incident photons per second per square meter, for each wavelength, as calculated from the 1.5 AM solar spectrum [46]. The total number of photons was the integrated number of photons absorbed in the 350-700 nm wavelength range (for P3HT:PCBM with permeable encapsulation) or 300-700 nm (for P3HT:PCBM with high-quality encapsulation).

FTIR spectra were measured in reflection mode due to the IR absorption of the glass substrate. The use of IR-transparent substrates was ruled out in order to assure heat conduction in the sample as similar as possible to that in common P3HT:PCBM solar cell devices, where the blend is deposited on ITO-coated glass substrates. Measurement were performed using a Thermo

Scientific Nicolet 8700 FTIR spectrometer, including a Smart Specul ATR accessory with a Ge crystal, a DTGS detector and a KBr beamsplitter, optical velocity: 0.3165, aperture: 138, 100 scans, 4 cm$^{-1}$ resolution. The samples were placed face-down on the Ge crystal, and pressed (180 – 200 N) to get good contact with the crystal.

Raman spectroscopy was carried out with a system comprised of a Jobin-Yvon LabRam HR 800 micro-Raman system, equipped with a liquid-N$_2$-cooled detector. The excitation source was an Argon laser (514.5 nm, 5 mW). For most measurements, ND filters were used to reduce the laser power to 0.005 mW, to prevent sample damage. The laser was focused with an x50 long-focal-length objective to a spot of about 6 μm diameter. Photoluminescence (PL) was measured using the same system at 520-804 nm range. The spectra were normalized by equalizing the C-C Raman peak at 1380 cm$^{-1}$ to account for thickness variations between samples.

## 3. Results and discussion

### 3.1. Sample temperature control

Figure 1 shows the effect of sunlight concentration on the temperature of encapsulated P3HT:PCBM blends with various temperature control schemes: attachment to a heat sink set to 5 or 25 $^o$C, IR light filtering (Fig. S4) and control via light chopping (Fig. S5). The results show a significant increase in the sample temperature with increasing irradiation in samples attached to a heat sink only without sunlight chopping or filtering, indicating that coupling the sample to a heat sink cannot fully control its temperature even at low sunlight concentrations. The temperature change rate under such conditions was found to be 0.2+/-0.01 $^o$C/sun (Fig. 1).

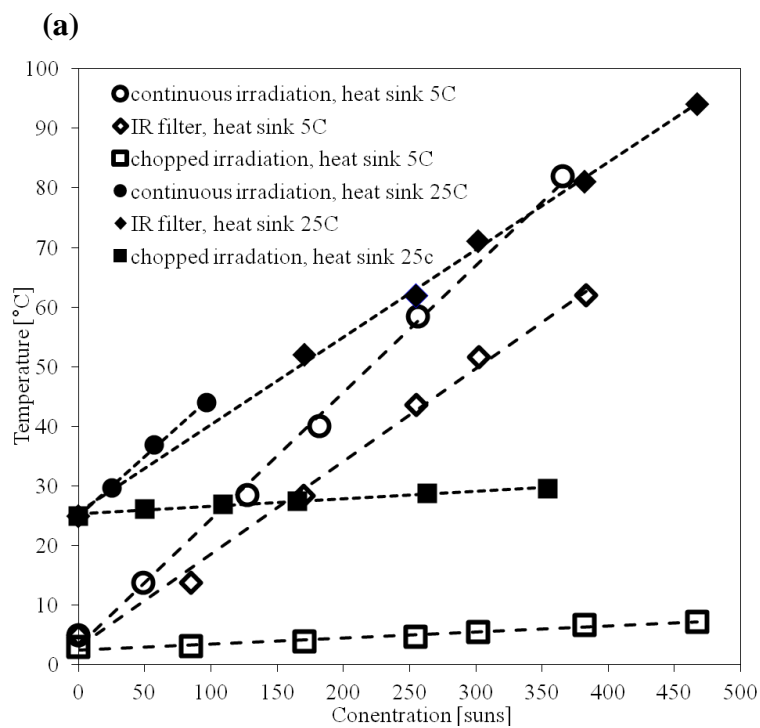

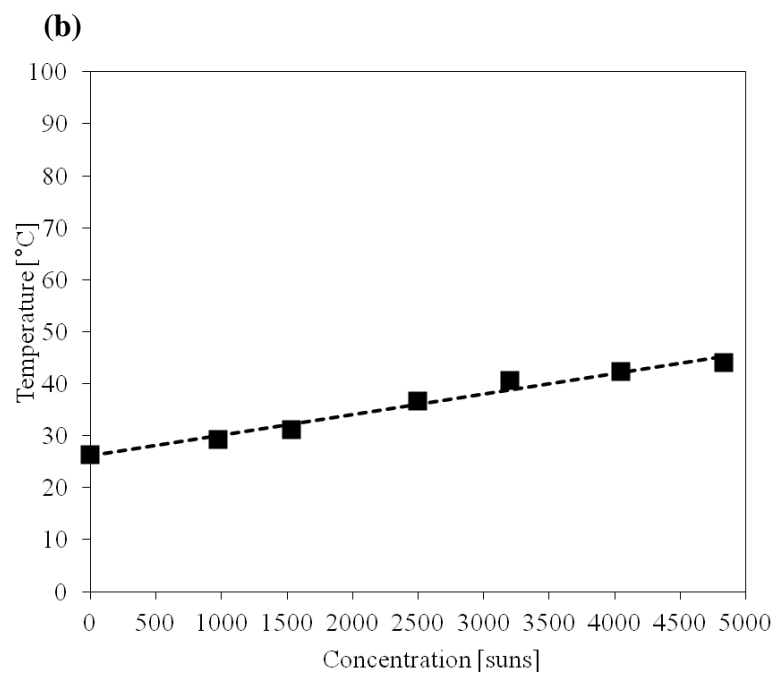

**Figure 1.** (a) Light intensity dependence of the sample temperature using various temperature control schemes as indicated in the legend, with linear data fits (dashed lines). The samples were illuminated in the 'solar furnace' through a hexagonal 0.55cm$^2$ kaleidoscope. (b) Light intensity dependence of the sample temperature under chopped irradiation using 0.0625 mm$^2$ kaleidoscope (higher concentration, heat sink at 25 $^o$C).

The linear dependence of the temperature on the irradiation intensity is expected from simple heat conduction principles, where at thermal equilibrium the heat gained by absorbed sunlight radiation equals the heat lost by conduction to the heat sink and convection in air above the sample (neglecting radiative heat losses):

$$P_{in} = P_{out} = P_{conduction} + P_{convection} = (k/d)A \cdot (T_{sample} - T_{heat-sink}) + h \cdot A(T_{sample} - T_{ambient}) \quad (1)$$

where $T_{sample}$, $T_{heat-sink}$ and $T_{ambient}$ are the temperatures of the cell, the heat-sink and the ambient, respectively, $A$ is the illuminated area, $k$ is the heat conductivity, $d$ is the sample thickness and $h$ is the convective heat transfer coefficient. Rearranging eq. 1 shows a linear dependence of $T_{sample}$ on $P_{in}$, the irradiating flux:

$$T_{sample} = \frac{P_{in}}{A(k/d + h)} + \frac{(k/d)T_{heat-sink} + h \cdot T_{ambient}}{(k/d + h)} \quad (2)$$

Adding the IR filter slightly reduced the sample heating rate to 0.15+/-0.005 °C/sun (Fig. 1), which can be explained by the ca. 30% reduction in illumination intensity ($P_{in}$) in the blend's absorption range due to the presence of the filter (Fig. S4). Clearly, sample heating was not eliminated by filtering the IR part of the spectrum. It can therefore be concluded that heating of P3HT:PCBM under concentrated sunlight is caused by photons at the absorbed wavelength range and dissipation of excess photon energy, and not necessarily by IR photon absorption.

In contrast, sunlight chopping was found to be an effective method for temperature control, with a negligible temperature change rate of ca. 0.01+/-0.005 °C/sun. The 95% reduction in the temperature change rate corresponds to the calculated irradiation time during each chopping cycle (5.4%, see section 2.2). We attribute it to heat dissipation during the 'light off' period in each cycle, preventing heat accumulation in the P3HT:PCBM. This behaviour is in agreement with our previously published measurements of light intensity dependence of $V_{oc}$ of P3HT:PCBM OPV cells, demonstrating no cell heating by chopped sunlight concentrated up to 100 suns [47]. Such behaviour is expected in high frequency chopping where the thermal inertia of the system smoothes temporal temperature fluctuations. We conclude that light chopping allows control of the sample temperature independent of the illuminating flux.

**3.2. Degradation of the P3HT:PCBM blend under concentrated sunlight with permeable encapsulation**

Figure 2 summarizes the results of degradation of light absorption of P3HT:PCBM samples with permeable encapsulation, allowing slow penetration of ambient species into the sample (see section 2.3). This allowed the study of photo-oxidation processes in P3HT:PCBM and the effect of illumination intensity and heating on them, towards comparison to such degradation in P3HT only [14]. The results were also compared to those following exposure in the absence of ambient species (well encapsulated cells, see section 3.3). The degradation in the absorption was quantified by the "degradation state" [14], corresponding to the degree of bleaching, calculated as the percent ratio of the total number of photons absorbed after sunlight exposure to that absorbed before exposure (see section 2.4).

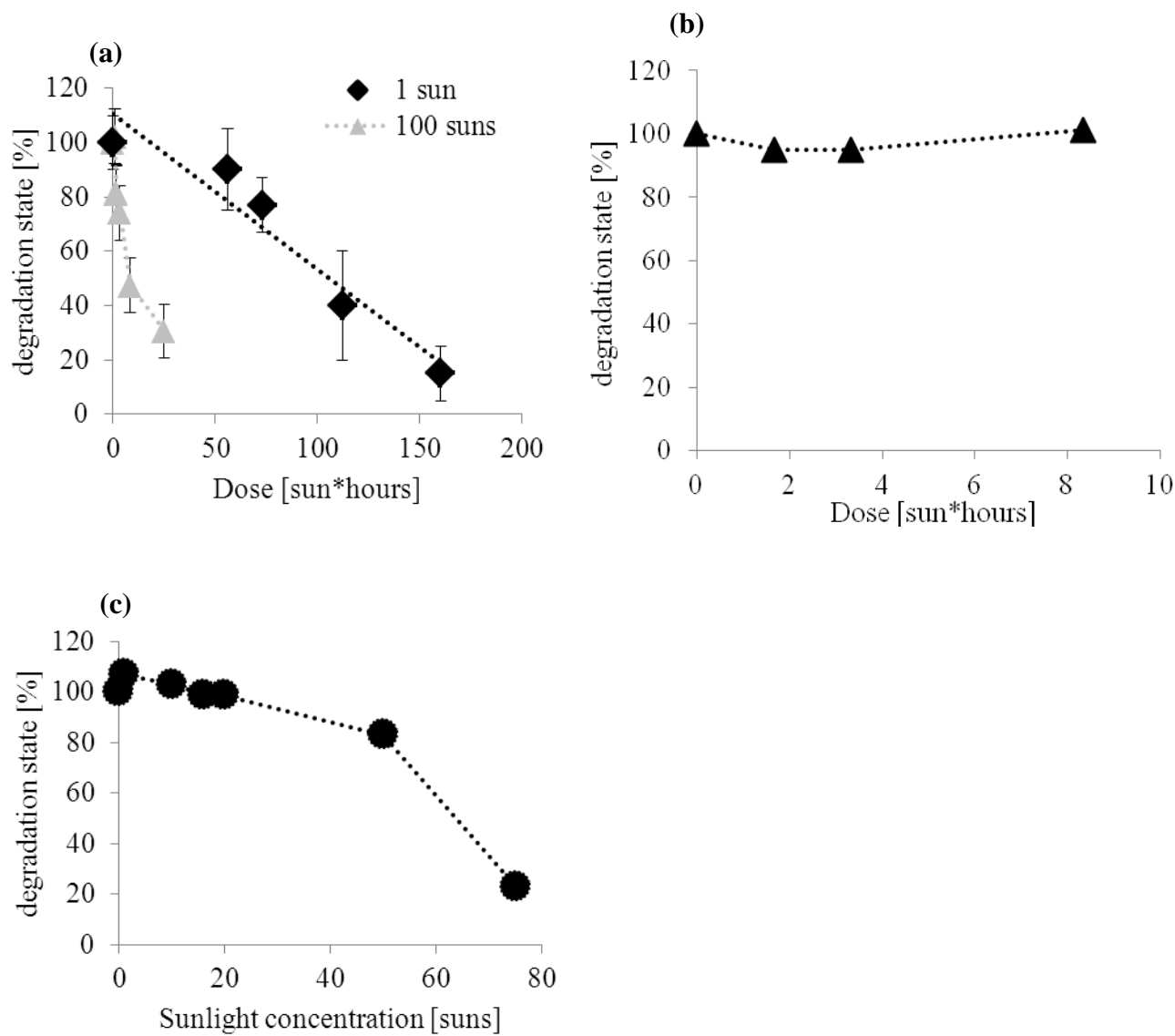

**Figure 2.** Absorption degradation of P3HT:PCBM films with permeable encapsulation as a function of dose of (a) continuous 1 and 100 suns, and (b) 100 suns chopped illumination, and (c) as a function of sunlight concentration for 1 hour of continuous exposure.

Figure 2a shows the absorption degradation state as a function of the illuminating dose for different sun concentrations under continuous illumination. The absorption degraded linearly with time at 1 sun, but higher concentration of 100 suns resulted in significantly accelerated degradation. The degradation rate at 100 suns was not constant and decreased with increased exposure time. T50 was reached after ca. 5 min. under 100 suns, compared to 100 hours under 1 sun, i.e., an acceleration factor of 1200. This behavior was attributed to the combined effect of high photon flux and heating under concentrated sunlight, which accelerated the thermally activated P3HT photo-oxidation, in agreement with such measurements of films of pure P3HT [14]. However, the degradation rate was slower than that of P3HT only, as expected, with complete degradation of the P3HT:PCBM blends' absorption under 1 sun after more than 190 hours compared to 50 hours for P3HT films [14, 16]. In contrast, chopped 100 suns illumination resulted in negligible degradation (less than 5%) up to 10 sun*hours, compared to almost 50% degradation after continuous exposure to a similar dose (Fig. 2b). This is yet another indication of the thermally activated nature of P3HT photo-oxidation [14]: when sample heating was minimized by light chopping, as shown above, the degradation rate was significantly slowed down, and photo-oxidation at low temperatures was relatively inefficient. For similar reasons, the photo-oxidation degradation rate increased super-linearly with sunlight concentration when the sample temperature was not controlled by chopping (Fig. 2c).

Detailed understanding of the degradation mechanism(s) under concentrated sunlight is needed for reliable accelerated stability testing of OPV devices. Increased light intensity may not only accelerate the degradation processes occurring at 1 sun exposure, but also stimulate "hidden" degradation mechanisms, as was previously shown for inverted OPV devices [17]. Therefore, one of the aims of the present study was to compare degradation processes under concentrated sunlight to those induced by exposure to 1 sun, to determine the concentration level and specific conditions at which the mechanism(s) are unchanged, validating the accelerated testing. The degradation mechanisms were followed using UV-vis and IR absorption, photoluminescence (PL) and Raman scattering spectroscopy.

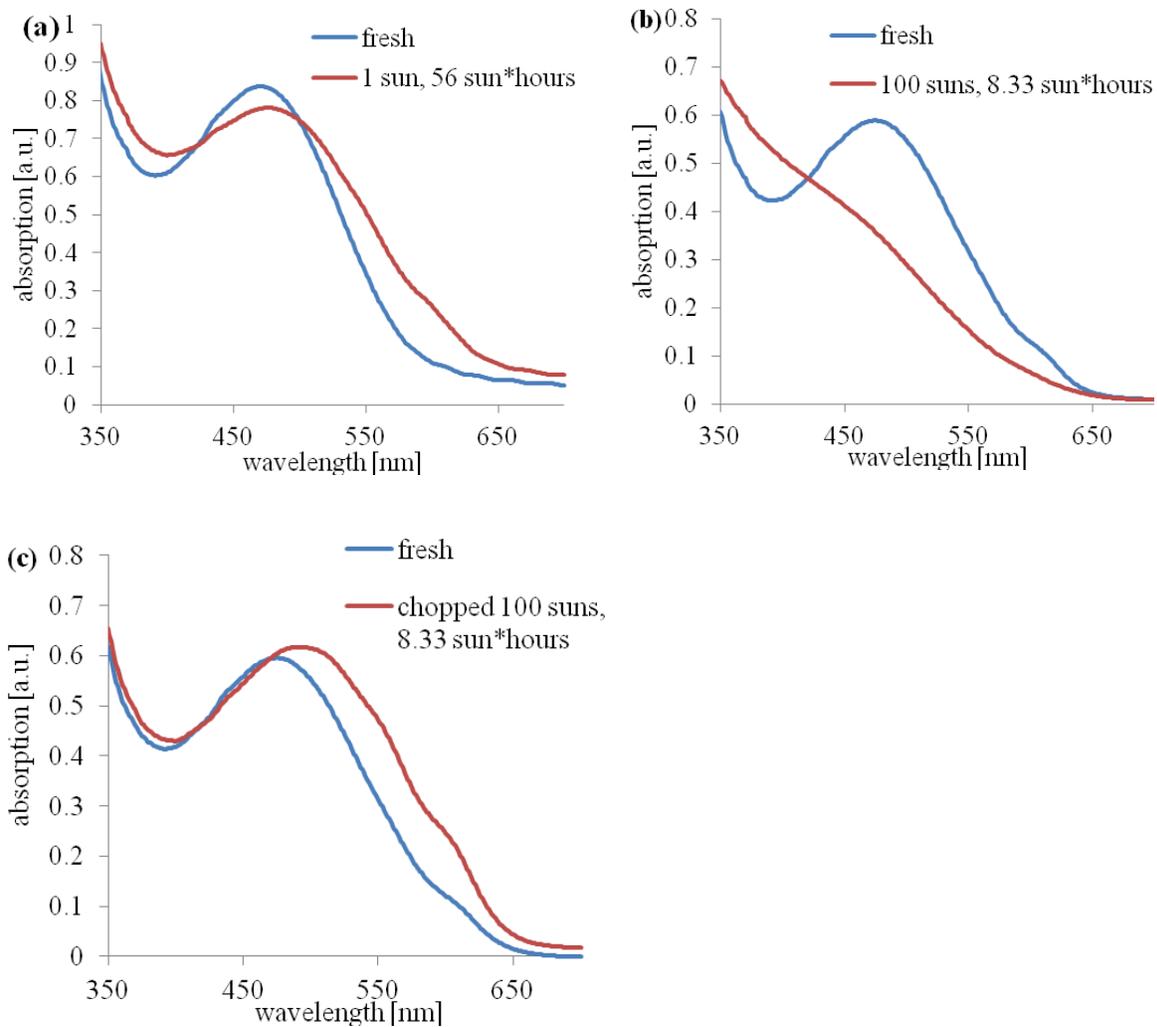

**Figure 3.** UV-vis absorption spectra of P3HT:PCBM films with permeable encapsulation under various exposures, compared to pre-exposure absorption of the same samples: (a) 1 sun, 56 sun*hours, (b) continuous 100 suns exposure, 8.33 sun*hours, (c) chopped 100 suns exposure, 8.33 sun*hours.

Figure 3 shows UV-vis absorption spectra of samples following various exposures to concentrated sunlight, which particularly demonstrate the temperature effect. 56 sun*hours of continuous exposure at 1 sun resulted in a decrease in the main (disordered) P3HT absorption peak at 470 nm (Fig. 3a). The absorption increase at longer wavelengths resulted from enhanced P3HT ordering and crystallization, identified by the shoulders at 554 and 605 nm and a shift of the main peak to 510 nm [48]. Continuous 100 suns exposure of a much smaller dose (8.33 sun*hours), but at a higher sample temperature (Fig. 1a), resulted in degradation in P3HT absorption over its entire absorption range (Fig. 3b). On the other hand, no absorption degradation was noted for the sample exposed to the same dose of *chopped* 100 suns, where the sample's temperature was kept low (Fig. 3c); rather, an increase in absorption at wavelengths longer than 470 nm is noted, attributed to an increase in P3HT chain ordering. No significant degradation attributed to PCBM absorption was detected.

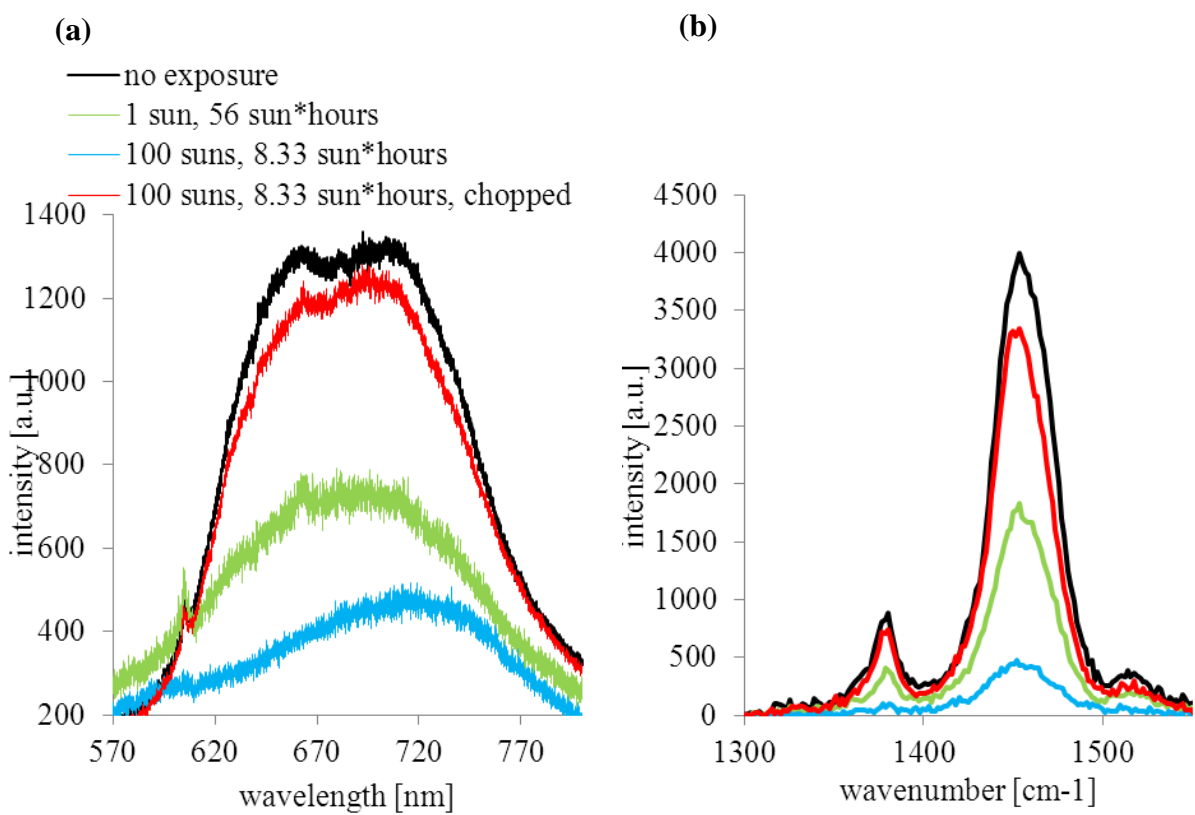

**Figure 4.** Photoluminescence (a) and Raman scattering (b) spectra of P3HT:PCBM samples after various exposures as indicated in the legend.

The PL emission spectrum of the un-exposed blend showed the expected peaks at 660 and 700 nm (Fig. 4a), related to the pure electronic transition and the first vibronic band in P3HT, respectively [49], with contribution from PCBM emission to the longer wavelength peak [22]. The relative decrease in PL intensity of the different samples is comparable to the relative decrease in the UV-vis absorption (Fig. 3). The largest decrease in P3HT emission, with a slight red shift, was noted after continuous exposure to 100 suns for 8.33 sun*hours, when the sample's temperature was significantly increased. The apparent red shift can be explained by the PCBM emission near 700 nm, while the P3HT emission is severely decreased, in agreement with the post-exposure absorption spectra (Fig. 3b) and the disappearance of ring-related IR absorption (Fig. S6). Exposure to 1 sun for 56 sun*hours also caused a significant decrease in the PL emission intensity, in accordance with the decrease in P3HT-related absorption. No significant PL degradation was noted for the sample exposed to chopped 100 suns, where the sample's temperature was kept low and there was no decrease in the absorption. The correlation with the UV-vis light absorption degradation points to the decrease in P3HT absorption as the main culprit for PL emission decay.

The Raman scattering spectra (Fig. 4b) show the expected P3HT ring-related peaks at 1380 and 1455 cm$^{-1}$, assigned to C-C skeletal stretching and C=C symmetric stretching, respectively, with a smaller peak at 1515 cm$^{-1}$ related to the C=C antisymmetric stretch mode [49-51]. The intensity of all peaks decreased with exposure to sunlight, indicating P3HT ring degradation. The relative decrease in intensity of the different samples is again comparable to the relative extent of P3HT degradation noted by other methods (UV-vis absorption, PL, and IR, Figs. 3, 4a, S6). This further confirms that preventing sample heating by chopping concentrated sunlight significantly reduced the degradation. Normalizing the spectra to the 1380 cm$^{-1}$ peak height (Fig. S7) showed that the ratio between the C-C and C=C related peaks as well as the general peak shape and width were conserved after degradation in all samples, hence no decrease in the conjugation length could be observed. We explain this by degradation in the quantity of the conjugated materials (indicated by the decrease in Raman emission intensity) but no change in the conjugation length of non-degraded chains (indicated by the conserved peak shape). This may indicate fast and abrupt P3HT degradation at certain areas more vulnerable to photo-oxidation, most likely the un-ordered P3HT domains, while less vulnerable areas, such as crystalline P3HT domains, were conserved. A closer look at the spectrum after the most severe degradation showed indication for the formation of new chemical species, showing as "shoulders" on the 1455 cm$^{-1}$ peak (Fig. S7), which are probably related to oxidized ring species.

The IR spectrum of the as-deposited P3HT:PCBM blend shows the expected absorption peaks, while exposure to sunlight in the presence of oxygen and humidity clearly induces fast degradation expressed in a decrease in peak intensities over the entire spectrum (Fig. S6). Decreased intensities of all typical absorption peaks points to P3HT ring scission and side chain oxidation as well as degradation of the PCBM's side group. However, the very small signal-to-noise ratio after degradation made a detailed comparison between differently exposed samples problematic. The overall similarity in the IR absorption spectra of the exposed samples indicates similarity in degradation products, i.e., in the degradation mechanisms.

Perusal of our results shows that the permeable encapsulation allows slow oxygen/moisture penetration into the samples, and that the observed degradation is governed by photo-oxidation of the P3HT backbone within the P3HT:PCBM blend. The process is significantly thermally accelerated, in agreement with previous observations of photo-oxidative degradation in P3HT-only films [14, 52]. We also postulate, based on the qualitative similarity of the spectroscopic results, that the degradation mechanisms are similar at one-sun and 100-suns exposures albeit at different rates. Similar degradation mechanisms validate the exposure to concentrated sunlight as a tool for accelerated aging tests of P3HT:PCBM blend, towards accelerated cell aging tests. Chopping concentrated sunlight allowed controlling the samples temperature, and demonstrated that P3HT photo-oxidation at room-temperature in P3HT:PCBM blends is rather slow.

**3.3. Exposure of well-encapsulated P3HT:PCBM blends to concentrated sunlight**

Concentrated sunlight exposure of well-encapsulated P3HT:PCBM samples did *not* cause significant degradation in light absorption (Fig. 5). This observation is in accordance with previous data about long term stability of well-encapsulated P3HT:PCBM cell under one sun exposure [1, 25, 53]. We demonstrated stability (~100% absorption) up to 3600 sun*hours of solar illumination (Fig. 5), a dose corresponding to ~1.6 years of operation at AM1.5G [54]. However, unlike previous stability tests at 1-sun exposure [1, 30], this dose was achieved after merely 12 hours of illumination by concentrated sunlight with intensity of 300 suns. A dose of 1200 sun*hours was applied by 15 min. of chopped 4800 suns concentrated sunlight illumination, i.e., total measurement time of 278 min. at well-controlled sample temperature. This demonstrates the significant advantage of concentrated sunlight for accelerated stability measurements.

Samples with additional ITO and PEDOT:PSS layers introduced between the P3HT:PCBM blend and glass substrate demonstrated similar behavior under concentrated sunlight as samples of P3HT:PCBM on glass (not shown), indicating that ITO and PEDOT:PSS layers do not accelerate the degradation of light absorption of the P3HT:PCBM active layer. This result is in agreement with previous data on stability in absorption upon exposure of P3HT:PCBM samples to UV-vis illumination in the absence of oxygen [55], however, a much larger irradiation dose was used herein.

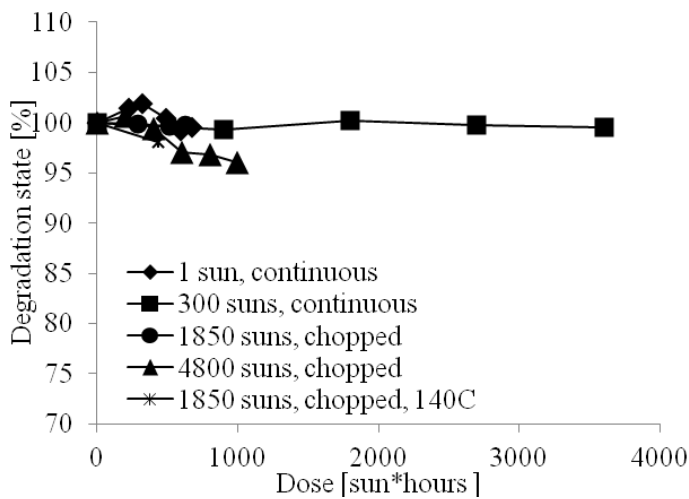

**Figure 5:** Absorption degradation state as a function of the illuminating dose for well-encapsulated P3HT:PCBM film after exposures as indicated in the legend.

Very small degradation in absorption was detected in P3HT:PCBM samples whose temperature was increased during the measurements (Fig. 5), i.e., samples either exposed to a very high sunlight concentration (chopped 4800 suns) or intentionally heated (chopped 1850 suns, 140°C). However, this degradation was limited to less than 5% after 1200 sun*hours, indicating a subtle degradation mechanism which is weakly thermally activated. Experiments with a higher dose (up to 3600 sun*hours achieved by 12 hours of continuous illumination) provided very similar results (Fig. S8). Figure 6 demonstrates the absorption spectra evolution of the well-encapsulated P3HT:PCBM film exposed to *chopped* illumination of 4800 suns (maximum dose of 1200 sun*hours) at controlled sample temperature. No degradation of P3HT-related absorption (i.e., in the main absorption peak at 510 nm and the ~554 and ~605nm shoulders, indicating ordering of P3HT domains in these pre-annealed samples [48]) was observed, hence even the weakly thermally activated degradation was eliminated by the temperature control induced by sunlight chopping. Slight degradation of PCBM-related absorption peak (345 nm) and absorption increase in the range 360-450 nm were noted. Due to the demonstrated stability of the well-encapsulated samples, no chemical analysis of the post-exposure samples was performed.

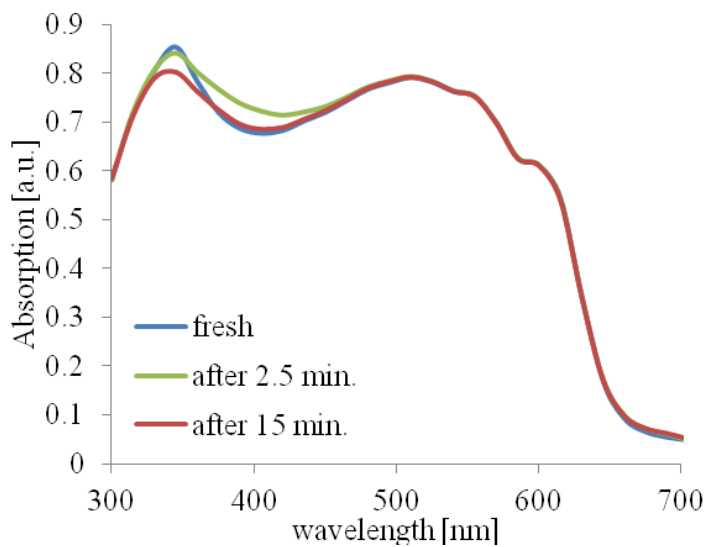

**Figure 6.** UV-vis spectra of well-encapsulated P3HT:PCBM blend samples exposed to chopped 4800 suns illumination.

These results demonstrate the high photochemical stability of P3HT in the P3HT:PCBM blend, where pure photolysis of P3HT in the absence of oxygen occurs at very low rate [22] and the presence of PCBM in the blend further suppresses it [55]. It also further confirmed our conjecture about absence of oxygen/moister penetration into the sample with this encapsulation method. Comparison of the degradation under continuous and chopped illumination *at similar doses* (Fig. 5) indicates differences that were attributed to the sample heating at continuous exposure, which was eliminated by chopping. Further work is needed to determine if this is the only factor causing the difference, or if processes occurring during the dark periods of the chopping cycles also affect the stability/ degradation.

## 4. Summary, Conclusions and Outlook

We have demonstrated OPV accelerated degradation studies using concentrated sunlight, where the atmosphere, temperature and illumination intensity were *independently* controlled. Testing various schemes for controlling the sample temperature under concentrated sunlight showed that heating of P3HT:PCBM was caused by photons at the absorbed wavelength range and dissipation of excess photon energy, and not necessarily by IR photon absorption. Therefore, IR filtering does not efficiently reduce the sample heating rate. Sunlight chopping was found to be an effective method for independent temperature control under illumination by concentrated sunlight. Various encapsulation methods demonstrated control of the atmosphere accessible by the active layer. The results demonstrated the ability of our studies to separate light- from heating- effects on degradation, and show the effect of these factors on different degradation mechanisms (photo-oxidation and photolysis).

The first accelerated degradation tests using sunlight concentration applied to P3HT:PCBM *blends* were reported. Various spectroscopic methods were utilized to determine the degradation and its mechanism(s). P3HT:PCBM blends exposed to concentrated sunlight in the presence of oxygen/ humidity showed degradation induced by photo-oxidation of the P3HT backbone within the P3HT:PCBM blend, which is significantly thermally accelerated, in agreement with previous observations. Chopping concentrated sunlight allowed controlling the samples temperature, showing that photo-induced oxidation at room-temperature is slow. Combined heating and concentrated sunlight illumination allowed significant acceleration rates for stability testing. Exposure of well encapsulated P3HT:PCBM films to concentrated sunlight demonstrated stability up to 3,600 sun*hours, corresponding to about at least 1.6 years of operating time. This result was obtained at 300 suns exposure after merely 12 hours, demonstrating the advantage of using concentrated sunlight for accelerated stability experiments.

The degradation/ stability tests using concentrated sunlight described herein can therefore combine extremely high acceleration factors with profound understanding of the effect of various, independently controlled factors on the degradation mechanisms. Independent control of light intensity, temperature and exposure to ambient opens new possibilities for studying their effect on various degradation mechanisms, for example, by using elevated temperatures at different illumination doses, or various illumination intensities at similar doses. Furthermore, chopped irradiation at various chopping frequencies can help detecting reversible phenomena in degradation mechanisms and their kinetics, and study the effect of processes occurring in the dark that could be affecting the degradation rate. Thus, this accelerated testing method provides an experimental tool for deeper understanding of OPV degradation and rapid screening of new OPV materials and devices. It can also constitute a basis for the development of a scaling relation between the degradation under 1 sun and under concentrated sunlight via rapid systematic determination of the acceleration factors, towards cross-validated accelerated stability testing.

**Acknowledgements**

EAK, IVF and YG acknowledge the support of the European Commission's StableNextSol COST Action MP1307. This work was also partially performed within the framework of the "Largecells" project funded by the European Commission's Seventh Framework Programme (FP7/2007-2014).

**Concentrated sunlight for accelerated stability testing of organic photovoltaic materials: Towards decoupling light intensity and temperature**

I. Visoly-Fisher, A. Mescheloff, M. Gabay, C. Bounioux, L. Zeiri, M. Sansotera, A. E. Goryachev, A. Braun, Y. Galagan, E.A. Katz

**Supplemental information**

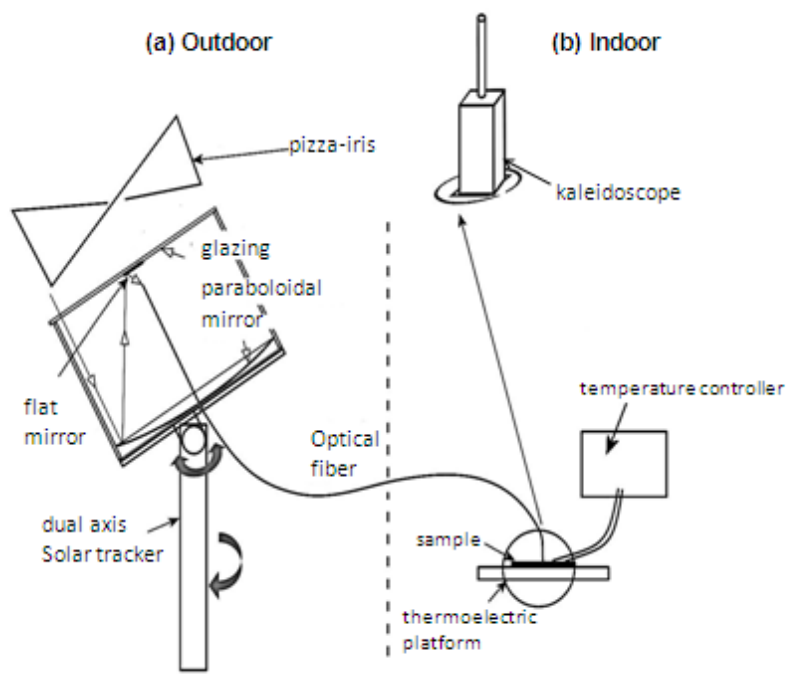

**Figure S1**. Schematic description of the Minidish dual-axis tracking solar concentrator (20 cm in diameter): (a) The outdoor set-up concentrates the solar radiation at the tip of a highly transmitting optical fiber, which guides the concentrated sunlight indoors onto the sample being tested, with (b) a uniform irradiation of the sample via a kaleidoscope [1].

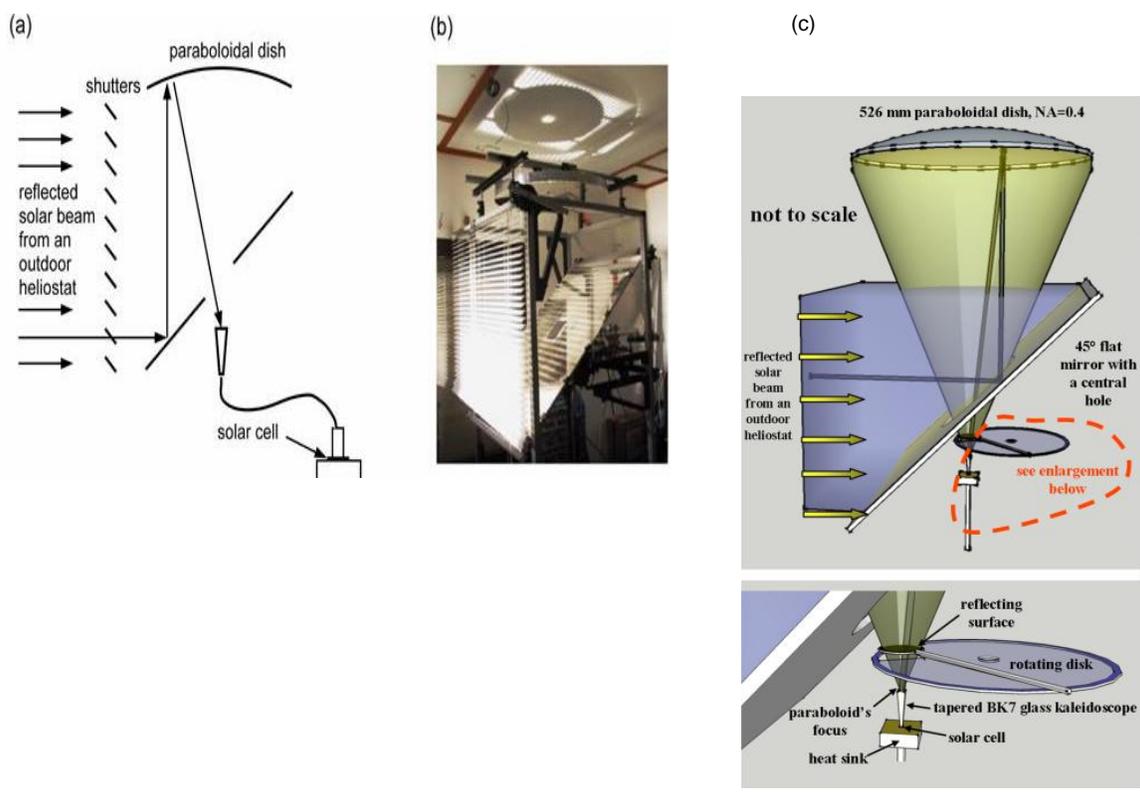

**Figure S2.** (a) Chopped-irradiation solar concentrator scheme: light from the heliostat passes the shutters and is reflected towards a paraboloidal, concentrated onto a tapered kaleidoscope and directed by an optical fiber toward a second kaleidoscope to the cell. (b) "Solar furnace" photo. (c) Schematic description of the concentrator operation in the flash-like mode [2].

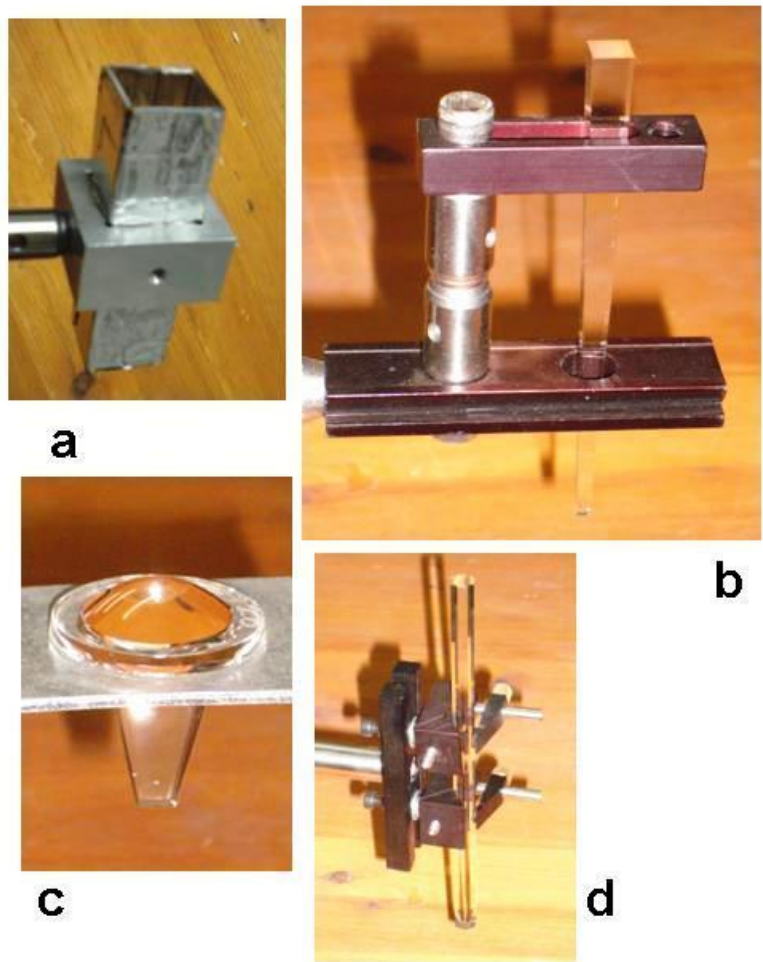

**Figure S3.** Kaleidoscopes used in the experiments: (a) 8 cm long mirror-based square kaleidoscope, exit area of 4 cm$^2$; (b) All-glass 10 cm long taped kaleidoscope, square exit area of 0.0625 cm$^2$; (c) All-glass 3 cm long taped kaleidoscope, square exit area of 0.3025 cm$^2$; (d) All-glass 15 cm long hexagonal kaleidoscope, exit area of 0.55 cm$^2$.

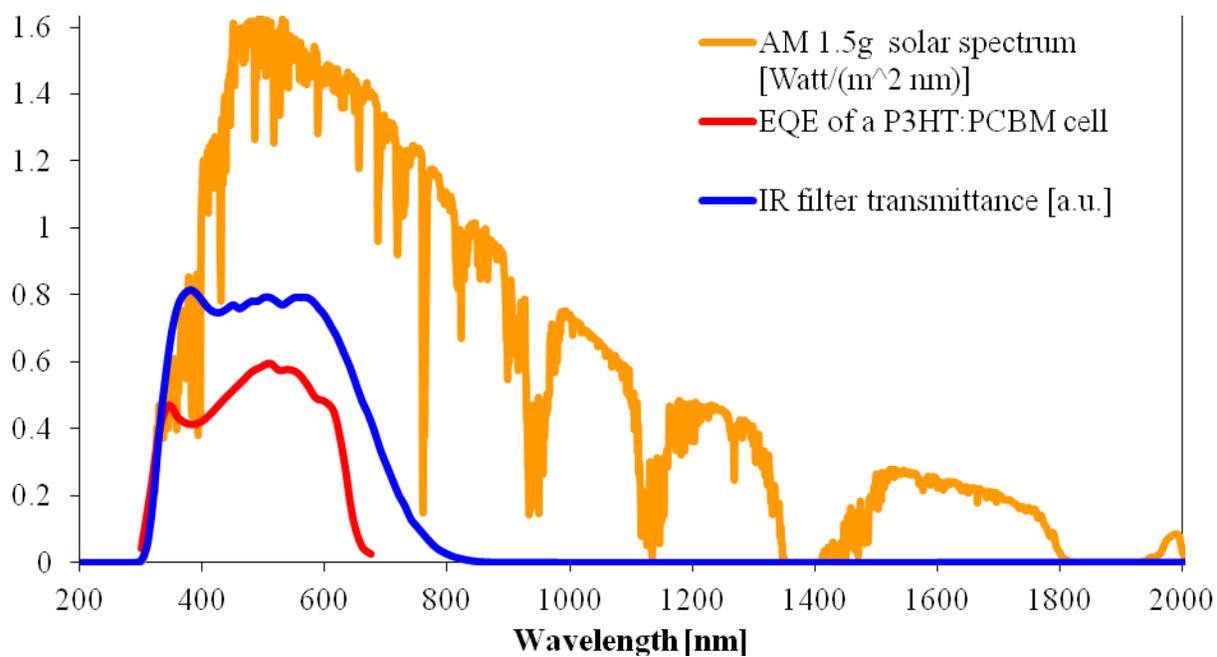

**Figure S4.** The transmittance spectrum of the IR short-pass filter. A typical EQE spectrum of a P3HT/PCBM cell and the AM 1.5g solar spectrum are shown for comparison.

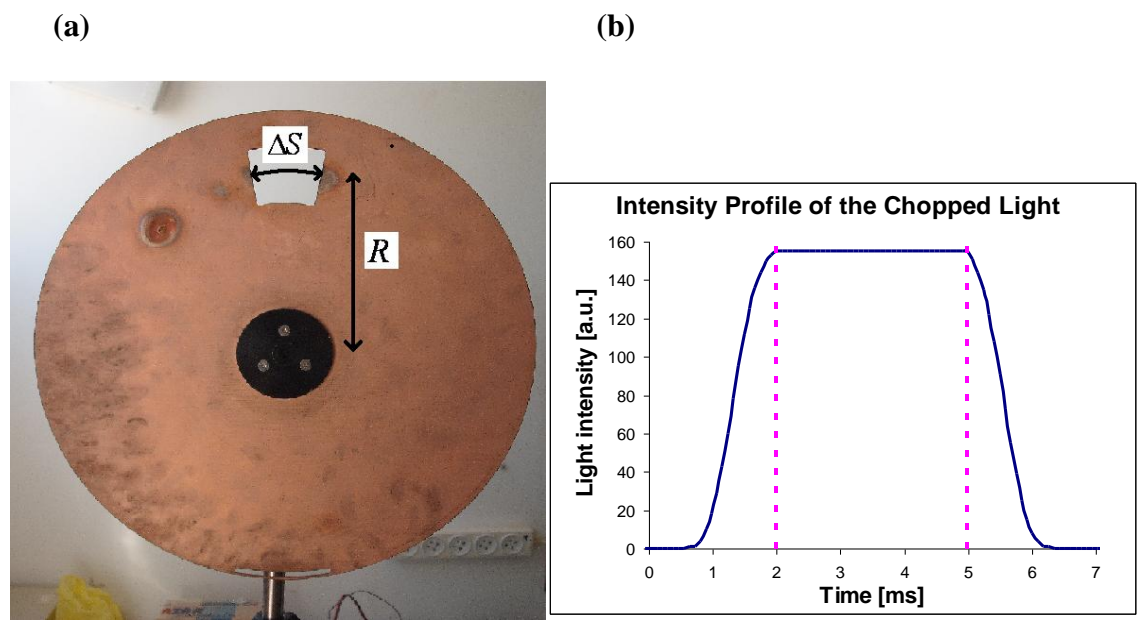

**Figure S5**. (a) Chopper disk used in the experiments. (b) Light intensity profile with the chopper rotating in our experiment with frequency of 13.5 Hz ($T = 74$ ms). Three temporal regions can be identified, starting with the sample in the dark: (i) opening (1-2 ms) - the sample is partially irradiated, (ii) constant irradiation (2-5 ms)- the sample is irradiated ($\tau=4$ msec) and (iii) closing (5-6 ms) )- the sample is partially irradiated. $<P_{in}>_{chopped}/<P_{in}>_{cont} = \tau/T = 0.054$.

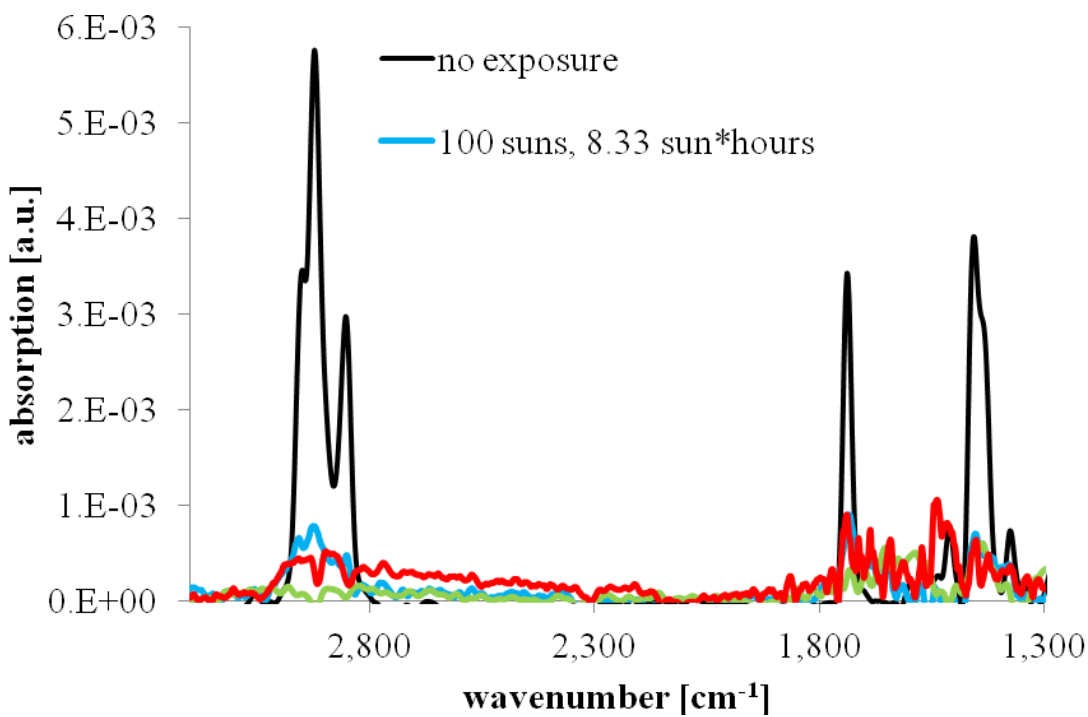

**Figure S6.** ATR-FTIR spectra of P3HT/PCBM samples with 'mild encapsulation' after various exposures as indicated in the legend. absorption peaks at 2955, 2925 and 2854 cm$^{-1}$ are typical of $CH_2$ and $CH_3$ stretch [3] in the P3HT alkyl side chains. The peak at 1740 cm$^{-1}$ results from C=O ester stretch in the PCBM [4, 5]. The 1510 cm$^{-1}$ peak is typical of the C=C stretch in the thiophene ring, as is also the 1454 cm$^{-1}$ absorption peak [3, 4, 6]. The 1510 cm$^{-1}$ and 1454 cm$^{-1}$ peaks are typical of the C=C stretch in the thiophene ring [3, 4, 6]. However, the large peak centered at 1450 cm$^{-1}$ also contains contributions from $CH_2$ and $CH_3$ stretches [3] and from the C60 vibrations in PCBM [7]. The peak at 1377 cm$^{-1}$ is also assigned to the decyl groups [3]

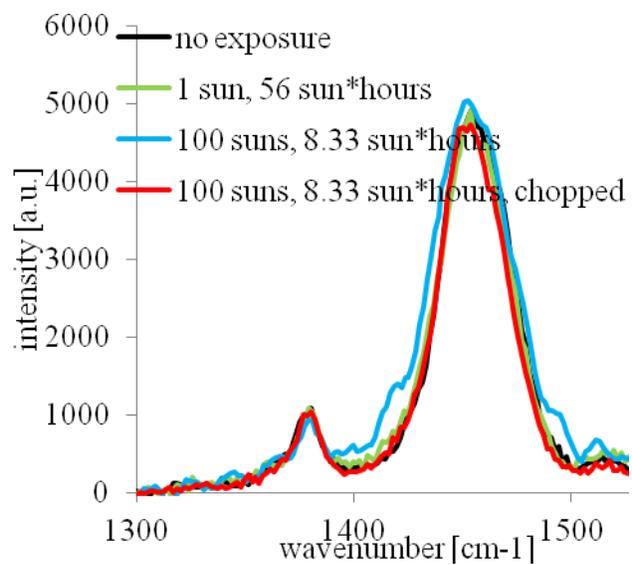

**Figure S7.** Raman scattering spectra of P3HT:PCBM samples after various exposures as indicated in the legend (normalized to similar 1380 cm$^{-1}$ peak heights).

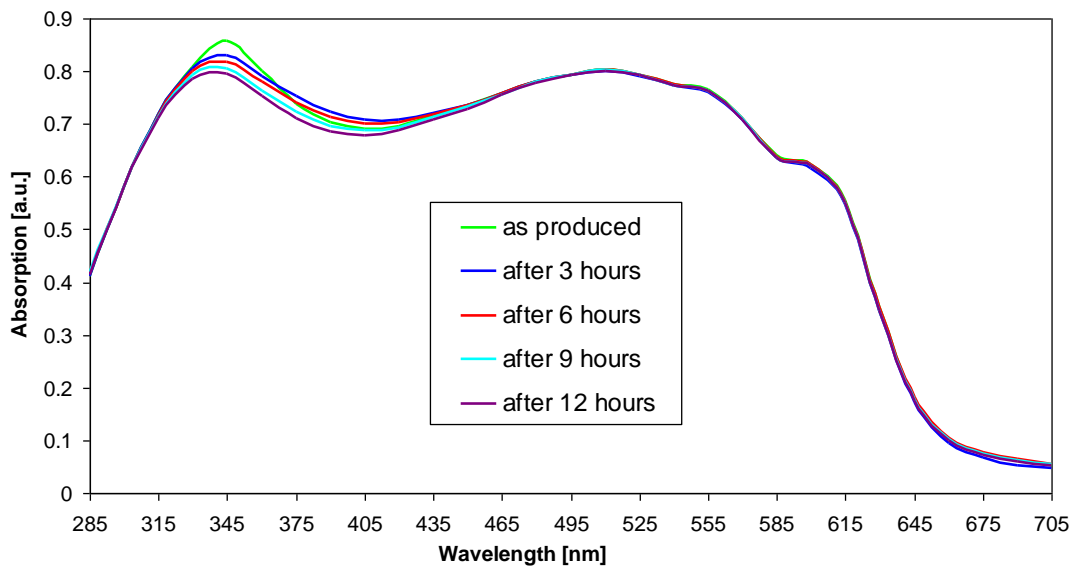

**Figure S8**. Light absorption spectra of the samples exposed to continuous, IR filtered irradiation of 300 suns.